# Electric-quadrupole and magnetic-dipole contributions to the $\nu_2+\nu_3$ band of carbon dioxide near 3.3 μm

Hélène Fleurbaey [1], Roberto Grilli [2], Didier Mondelain [1], Samir Kassi [1], Alain Campargue [1]*

[1] *Univ. Grenoble Alpes, CNRS, LIPhy, 38000 Grenoble, France*
[2] *Univ. Grenoble Alpes, CNRS, IRD, Grenoble INP, IGE, 38000 Grenoble, France*

The recent detections of electric-quadrupole (E2) transitions in water vapor and magnetic-dipole (M1) transitions in carbon dioxide have opened a new field in molecular spectroscopy. While in their present status, the spectroscopic databases provide only electric-dipole (E1) transitions for polyatomic molecules ($H_2O$, $CO_2$, $N_2O$, $CH_4$, $O_3$…), the possible impact of weak E2 and M1 bands to the modeling of the Earth and planetary atmospheres has to be addressed. This is especially important in the case of carbon dioxide for which E2 and M1 bands may be located in spectral windows of weak E1 absorption. In the present work, a high sensitivity absorption spectrum of $CO_2$ was recorded by Optical-Feedback-Cavity Enhanced Absorption Spectroscopy (OFCEAS) in the 3.3 μm transparency window of carbon dioxide. The studied spectral interval corresponds to the region where M1 transitions of the $\nu_2+\nu_3$ band of carbon dioxide were recently identified in the spectrum of the Martian atmosphere. Here, both M1 and E2 transitions of the $\nu_2+\nu_3$ band were detected by OFCEAS. Using recent *ab initio* calculations of the E2 spectrum of $^{12}C^{16}O_2$, intensity measurements of five M1 lines and three E2 lines allow us to disentangle the M1 and E2 contributions. Indeed, E2 intensity values (on the order of a few $10^{-29}$ cm/molecule) are found in reasonable agreement with *ab initio* calculations while the intensity of the M1 lines (including an E2 contribution) agree very well with recent very long path measurements by Fourier Transform spectroscopy. We thus conclude that both E2 and M1 transitions should be systematically incorporated in the $CO_2$ line list provided by spectroscopic databases.

\* Corresponding author: alain.campargue@univ-grenoble-alpes.fr

## 1. Introduction

The infrared absorption in the atmosphere of the Earth and other planets mainly arises from rovibrational absorption lines due to electric-dipole (E1) transitions listed in standard spectroscopic databases such as HITRAN [1] or GEISA [2]. E1 transitions result from the coupling of the electromagnetic field with the transition electric-dipole moment induced by molecular vibration. Currently, E1 rovibrational transitions are the only transitions included in databases for polyatomic molecules ($H_2O$, $CO_2$, $O_3$, $N_2O$, $CH_4$…) and heteronuclear (or polar) diatomic molecules (CO, HF...). In the case of homonuclear diatomic molecules ($H_2$, $N_2$, $O_2$...), the vibration does not induce an electric-dipole moment and E1 transitions have no intensity. Allowed, though exceptionally weak, quadrupolar transitions (E2) of these species have been detected and consequently included in spectroscopic databases. E2 lines are weaker than E1 lines by typically six orders of magnitude, making their detection particularly challenging in the laboratory [3-11]. Interestingly, up to recently, the HD isotopologue of hydrogen, which has a small dipole moment, was the only species for which both E1 and E2 transitions were measured (see review included in Ref. [11]). This applies now also to water vapor for which very weak E2 transitions were reported from high sensitivity laboratory spectra recorded by cavity ring down spectroscopy (CRDS) near 1.3 µm [12] and Fourier transform spectroscopy near 2.5 and 5.4 µm [13]. These detections were made possible by accurate *ab initio* predictions of the E2 transitions of $H_2^{16}O$ [12]. Indeed, in the case of water vapor, E2 vibrational bands are also activated by the E1 mechanism, hence E2 bands coincide with the much stronger E1 bands, even though, due to different rotational selection rules, E1 and E2 lines do not coincide in positions. Most of the E2 lines are thus hidden in the profile of the numerous and much more intense E1 lines, and accurate calculations are needed to identify narrow spectral intervals between E1 lines [12,13] suitable for the detection of E2 transitions.

In the case of carbon dioxide, the E2 spectrum has recently been computed theoretically [14]. On one hand, the detection of E2 lines in $CO_2$ appears to be more challenging than in water because the strongest $CO_2$ E2 lines ($7\times10^{-29}$ cm/molecule near 3.3 µm [14]) are more than two orders of magnitude weaker than the strongest water E2 lines ($10^{-26}$ cm/molecule near 2.5 µm). On the other hand, in $CO_2$, contrary to water vapor, the vibrational selection rules ensure that E2 bands are not activated by the E1 mechanism and thus may be located in spectral ranges where the E1 absorption is very weak (the $CO_2$ transparency windows) [12, 15].

The $CO_2$ windows are important for the detection of trace gases in planetary atmospheres, for instance those of Venus and Mars, where carbon dioxide is the major atmospheric absorber. The 3.3 µm window is particularly suitable to detect CH compounds like methane, which has its extremely strong $\nu_3$ band (line intensities larger than $10^{-19}$ cm/molecule) centered in this $CO_2$ window. In the frame of the ESA-Roscosmos ExoMars Trace Gas Orbiter (TGO) mission, a mid-infrared channel has been optimized to record very high sensitivity spectra of the Martian atmosphere near 3.3 µm (the achieved detection limit for methane was estimated to 0.05 ppbv) [16,17]. Although methane has not been detected so far, a series of very weak regularly spaced absorption lines were detected in the region and remained unexplained up to recently. This series of lines, absent in the available carbon dioxide databases, were very recently assigned to the ν$_2$+ν$_3$ band of $^{12}C^{16}O_2$ [17], which is forbidden in E1 absorption and predicted to be the strongest E2 band in $^{12}C^{16}O_2$ by *ab initio* calculations. Nevertheless, in spite of the agreement of line positions, the observed intensity distribution of the band ruled out the E2 ν$_2$+ν$_3$ assignment of the observed band. Specifically, the ExoMars band shows a strong *Q* branch while *Q* line transitions are predicted to be very weak for the E2 band. On the basis of theoretical considerations, including detailed consideration of the absorption selection rules [15], Trokhimovskiy et al. assigned the ExoMars band to the ν$_2$+ν$_3$ band activated by a magnetic dipole (M1) mechanism [17]. This is the first identification of M1 transitions between vibrational states of a molecule in the $X^1\Sigma$ ground electronic state which has neither spin nor electronic angular momentum. The existence of a magnetic dipole is due to the vibrational angular momentum ($l_2$= 1) resulting from the excitation of the doubly degenerate bending mode of the $(v_1 v_2{}^{l_2} v_3) = (01^1 1)$ upper level [15].

The aim of the present investigation was to record the $CO_2$ spectrum near 3.3 μm at high sensitivity in well-controlled laboratory conditions, in order to detect the E2 ν$_2$+ν$_3$ band predicted by theory. The first laboratory measurement of these E2 lines will allow us to test the quality of the *ab initio* intensity values of the E2 line intensities recently computed [14] and to disentangle the M1 and E2 contributions in the ν$_2$+ν$_3$ band of $^{12}C^{16}O_2$.

**2. Experiment**

*2.1. The OFCEAS spectrometer*

The recordings were made using the optical-feedback cavity enhanced absorption spectroscopy (OFCEAS) technique [18,19]. This cavity-enhanced absorption technique has been extensively used for trace gas detection, but its high sensitivity and reliability made it a good choice for the first detection of these very weak $CO_2$ lines.

The experimental setup is illustrated in **Fig. 1** and described in detail in Ref. [20]. A 3.3 μm Nanoplus interband cascade laser (ICL) is locked by optical feedback to a V-shaped high finesse cavity. The design of the cavity ensures that only photons in resonance with the cavity participate to the optical feedback, and not for instance photons reflected at the entrance [18,19], thus narrowing the laser emission and ensuring an optimum coupling efficiency. To satisfy the feedback phase matching, the laser is placed at a distance of exactly one cavity arm length (40 cm) from the entrance of the cavity. One of the stirring mirrors is mounted onto a piezoelectric transducer to finely tune the phase of the feedback and obtain a mode-by-mode locking. As the laser current is ramped, the laser frequency remains temporarily locked onto successive TEM$_{00}$ modes. A polarizer is used to adjust the feedback rate. As the laser current is tuned, the peak transmission of each cavity mode is recorded. It is then divided by a virtual reference signal representative of the expected laser intensity to current dependence. The inverse square root of this ratio, $H_{max}$, is directly proportional to the absorption coefficient $\alpha$ of the medium inside the cavity [19]: for each mode *m*, $\alpha(m) = \beta/\sqrt{H_{max}(m)}$. In order to calibrate the measurement, a single measurement of the exponential decay time (ring-down) of the photons inside the resonator is recorded at the end of each laser scan (on the mode k), with a ring down decay time $\tau(k) = 1/c\alpha(k)$, where $c$ is the speed of light. Then for each mode $m$, the absorption coefficient is derived as: $\alpha(m) = \frac{1}{c\tau(k)}\sqrt{\frac{H_{max}(k)}{H_{max}(m)}}$ [21].

In the present study, the cavity mirrors (produced by Layertec) have a measured reflectivity of 99.9914 ± 0.0003 % around 3.306 μm corresponding to a ring-down decay time of about 15.5 ± 0.5 μs. The V-shaped cavity has a total length of 80 cm leading to a free spectral range (FSR) of 187.4 MHz.

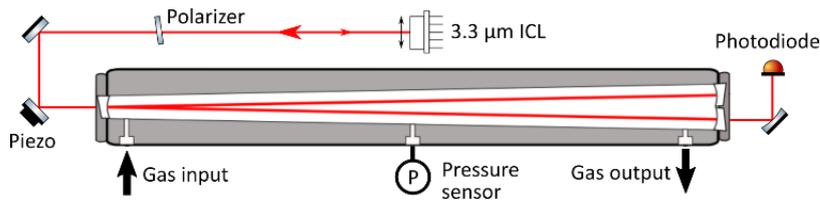

**Fig. 1.** Schematic diagram of the experimental setup. The polarizer allows for a fine control of the optical-feedback level. The Piezo permits to control the feedback phase. The signal of the photodiode is normalized with a virtual reference, representative of the incident laser power, which is not physically measured.

The gas sample was composed of pure $CO_2$ (AlphaGaz2 from Air Liquide, impurities < 0.1 ppm). It was introduced in the cell in a flowing configuration, with a flow controller (IQF, Bronkhorst) at the entrance of the cell and an electrovalve (FAS, Norgren) at the output to control the intra-cavity pressure. Spectra were recorded at a flow rate of about 9 sccm (standard cubic centimeter per minute). The pressure was measured by a Wika 0-250 mbar pressure gauge (Model S-10, accuracy ≤ ±0.50 % of span or ±1.25 mbar). Absorption lines of water and methane, present as impurities in the sample, were visible in the spectra. The water concentration was reduced by passing the gas through a copper tube coiled inside a Dewar filled with ice and water. The remaining water concentration (evaluated from the line at 3023.15 cm$^{-1}$ to be about 50 ppm) was attributed to outgassing from the walls of the cell, which has been extensively used with very high concentrations of water in previous experiments. The methane concentration in the sample was estimated from the 3028.74 cm$^{-1}$ line which has a very large intensity (9.21×10$^{-20}$ cm/molecule). Although this line is superimposed on the "M1" $CO_2$ line at 3028.75 cm$^{-1}$, we were able to estimate the $CH_4$ concentration to 4 ± 1 ppb. The temperature of the cell was stabilized at ~40°C (see next section) thanks to a resistive heating band, a PT1000 sensor and a PID regulation.

*2.2. Data acquisition*

The spectra recordings targeted a few E2 lines of the $\nu_2+\nu_3$ band predicted by *ab initio* calculations [14] between 3023 and 3031 cm$^{-1}$. For each scan recorded in 125 ms, the ICL current was tuned over 13 mA while the laser temperature was kept constant, resulting in a scan width of about 145 cavity modes or ~0.9 cm$^{-1}$.

Several laser scans were interlaced to further increase the resolution and average over spectral noise (the FSR is 0.00625 cm$^{-1}$ to be compared to a Doppler width of about 0.0029 cm$^{-1}$ (HWHM)). To this end, the temperature was slowly tuned during 10 minutes from 41°C to 39°C by switching off the heating element of the cavity, while about 1800 laser scans were recorded. The temperature change caused the cavity modes positions to slowly drift over time. In each scan, a spectral feature was fit with a Voigt profile to determine the frequency shift of the mode positions, and the scans were arranged in bins corresponding to 1/10 of the FSR and then averaged within each bin, effectively increasing the spectral resolution by a factor of 10 to 18.74 MHz while averaging about 180 measurements for each data point. **Figure 2** shows three individual scans (symbols and lines in various shades of grey) along with the spectrum (red line) resulting from the interlacing of 1800 scans.

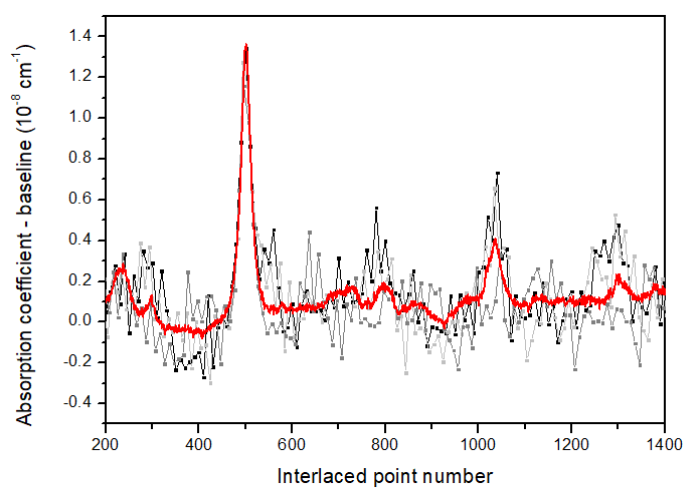

**Fig. 2.** Raw and interlaced spectra recorded around 3024 cm$^{-1}$ at 80 mbar. Three raw spectra (mode spacing of 1 FSR) are shown along with the spectrum resulting from the interlacing of 1800 spectra (red line, interlaced point spacing of 1/10 FSR = 18.74 MHz).

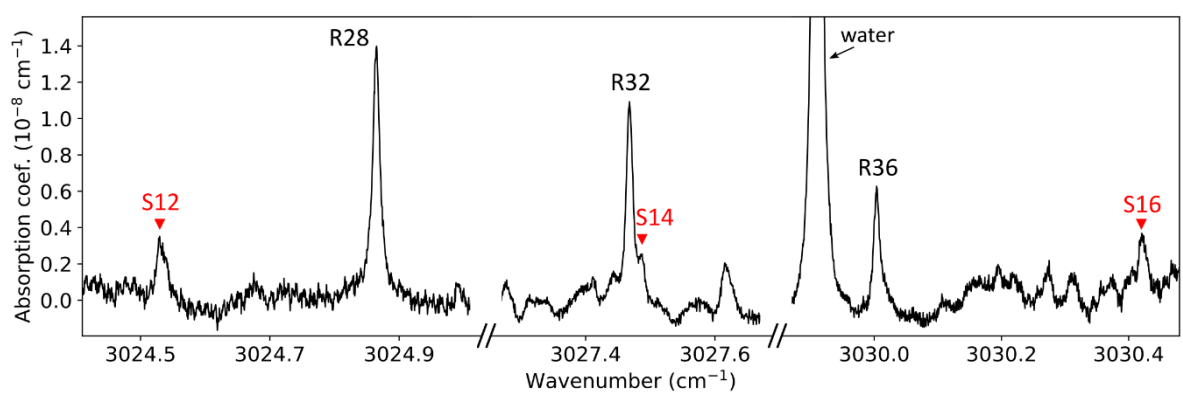

**Fig. 3.** Interlaced spectra in three spectral intervals, showing the three electric-quadrupole transitions observed in this work (S12, S14 and S16), along with three "M1" transitions (R28, R32 and R36). In the right-hand plot, a water line is visible at 3029.9 cm$^{-1}$. These spectra were acquired at a pressure of 60 mbar.

Spectra were recorded in five spectral intervals within the tuning range of our laser, encompassing five M1 transitions and three pure E2 transitions. For each wavelength region, spectra were acquired at three pressure values: 40, 60 and 80 mbar. In **Fig. 3**, three interlaced spectra recorded at a pressure of 60 mbar are plotted to present the three newly-observed electric-quadrupole lines.

*2.3. Analysis*

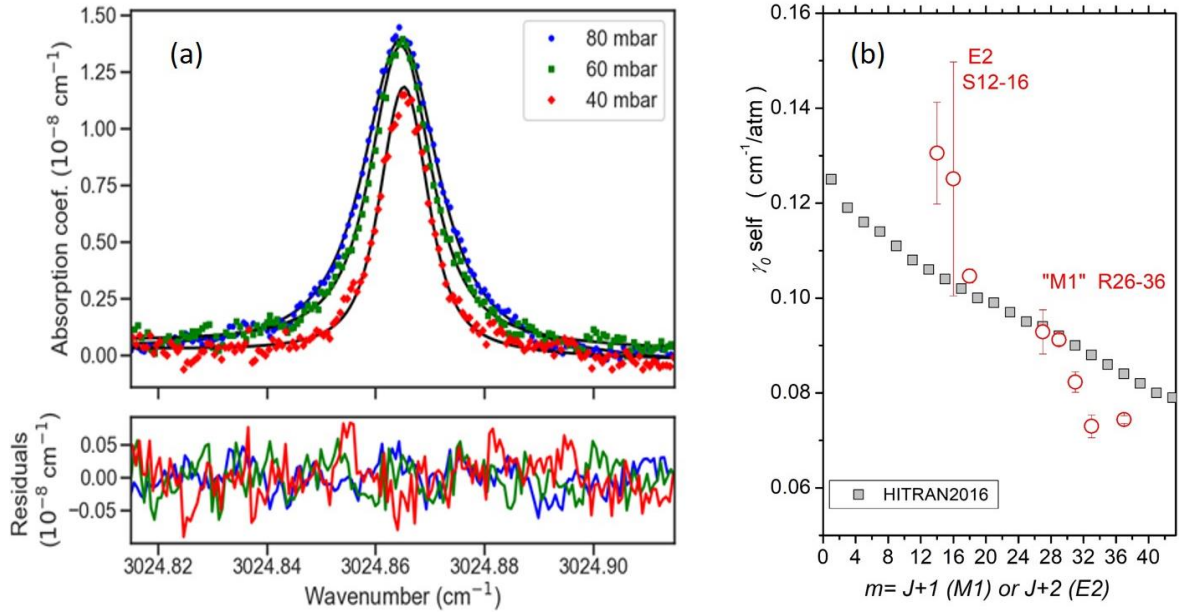

**Fig. 4**.
(a) Multi-spectrum analysis of the R28 line of the $\nu_2+\nu_3$ band of $^{12}C^{16}O_2$ recorded at 40, 60 and 80 mbar. The upper plot shows the interlaced spectra recorded for three pressure values (colored symbols) along with the fitted Voigt profiles (black lines). The lower plot shows the residuals of the fits (obs. – calc.).
(b) Self-broadening coefficients obtained through this multi-spectrum analysis, compared to the "universal" *m* empirical dependence from HITRAN2016.

A multi-spectrum analysis was performed for each line using the MATS (Multi-spectrum Analysis Tool for Spectroscopy) software developed at the National Institute of Standards and Technology [22]. Using this software written in the Python language, the lines were fit individually, using a set of three interlaced spectra (one for each pressure), with a model consisting of a Voigt profile and a baseline. The baseline was modeled by a linear function. For some of the spectra an optical fringe with a period of 1.6 GHz was visible, and it was included in the baseline by fitting its amplitude and phase. The Doppler width of the Voigt profile was fixed to the theoretical value calculated for the average temperature of 40°C, while the collisional broadening coefficient and intensity were floated but constrained to be independent of pressure. The line position was also added as a fit parameter since the frequency axis was not absolute but only empirically determined. The result of this fitting procedure for the R28 line at 3028.865 cm$^{-1}$ is shown in **Fig. 4a**. Because of the weak signal-to-noise ratio (ranging from 5 to 60 depending on the line) the fit results were sensitive to the baseline and the extent of the spectral range included in the fit. For this reason, the fitting procedure was repeated for several spectral ranges around each line (between 0.06 and 0.18 cm$^{-1}$ wide), and the average results are reported in **Table 1**. In the MATS software, the line intensity is automatically converted to the 296 K reference temperature according to the lower state energy value and partition functions provided by the HITRAN2016 database. Statistical error bars provided by the fit are given in **Table 1**. It remains difficult to estimate quantitatively the total uncertainty values. Residual fringes and the superposition with unidentified lines may have a significant contribution to the error budget (see **Fig. 3**). As a result, an estimated value of 8 % and 30 % seems reasonable for the total uncertainty of the "M1" and E2 line intensities.

In the considered 40-80 mbar range, the measured line profiles are significantly affected by pressure broadening (see Fig. 4, left-hand). The self-broadening coefficients derived from the multi-spectrum analysis are included in **Table 1** and plotted as a function of $m$ in **Fig. 4b** ($m = J+1$ and $J+2$ for $R(J)$ and $S(J)$ transitions, respectively). Although some values are affected by large uncertainties, a clear decrease of $\gamma_{self}$ with $m$ is noted, E2 lines corresponding to smaller $m$ values being wider that the observed "M1" lines. The obtained values roughly follow the "universal" $m$ dependence adopted for all the E1 bands in the HITRAN database. We thus conclude that in a first approximation the E1 $m$ dependence of $\gamma_{self}$ applies to M1 and E2 lines.

**Table 1.**
Parameters of the M1 and E2 absorption lines of the $\nu_2+\nu_3$ band of $^{12}C^{16}O_2$ measured by OFCEAS near 3.3 μm. The line positions are empirical values from Ref. [23].

| Position (cm$^{-1}$) | Transition | Type | Intensity $^a$ ($10^{-28}$ cm/molecule) | $\gamma_{self}$ $^b$ (cm$^{-1}$/atm) |
|---|---|---|---|---|
| 3023.5385 | R26 | M1+E2 | 1.926(74) | 0.0929(15) |
| 3024.5297 | S12 | E2 | 0.545(58) | 0.1305(107) |
| 3024.8652 | R28 | M1+E2 | 1.87(3) | 0.0912(11) |
| 3026.1751 | R30 | M1+E2 | 1.41(26) | 0.0823(22) |
| 3027.4684 | R32 | M1+E2 | 1.24(1) | 0.0730(24) |
| 3027.4872 | S14 | E2 | 0.322(36) | 0.1251(246) |
| 3030.0044 | R36 | M1+E2 | 0.772(8) | 0.0744(8) |
| 3030.4233 | S16 | E2 | 0.487(21) | 0.1047(42) |

$^a$ Statistical error bars provided by the fit are given within parenthesis in the unit of the last quoted digit. Note that the total uncertainties are larger (about 8 and 30 % for "M1+E2" and E2 transitions, respectively- see Text).
$^b$ Self-broadening coefficient (cm$^{-1}$/atm),

### 3. Discussion

The above identification of the M1 lines in our spectra was obtained by comparison with the empirical line list of the $\nu_2+\nu_3$ M1 band of $^{12}C^{16}O_2$ provided in Ref. [24]. In Ref. [15], Perevalov et al. presented the M1 (and E2) transition selection rules and derived the theoretical expression of the M1 line intensities including the Hönl-London factors and transition magnetic dipole moment (MDM) squared. In Ref. [24], intensity values of some lines of the $\nu_2+\nu_3$ M1 band in the $P$, $Q$ and $R$ branches were reported from FTS spectra of $CO_2$ recorded with a pathlength of 1058 m and pressures of 182, 291 and 380 mbar. The FTS line intensities were used to fit the vibrational transition MDM and generate a complete line list of the $\nu_2+\nu_3$ M1 band (see grey stars in **Fig. 5**). Line positions were obtained with an accuracy better than 0.001 cm$^{-1}$ from the energy levels calculated using the effective Hamiltonian parameters published by Majcherova et al. [23]. (This effective Hamiltonian is the same as used to generate the Carbon Dioxide Spectroscopic Databank [25] and most of the HITRAN energy levels and line positions of $^{12}C^{16}O_2$.)

As regards E2 lines in the OFCEAS spectra, their assignment relied on the theoretical E2 list with intensities derived in Ref. [14] from an *ab initio* quadrupole moment of $CO_2$ and line positions empirically adjusted according to Ref. [23] in the same way as described above for M1 transitions. The corresponding E2 line list (not limited to the $\nu_2+\nu_3$ band) is displayed in **Fig. 5** (open red stars). Let us mention that according to these *ab initio* predictions, the $\nu_2+\nu_3$ band is the strongest E2 band and the 3.3 µm window is the most favorable for an E2 detection in $CO_2$ [14].

Differences between M1 and E2 selection rules make it possible to detect pure E2 transitions (see Fig. 1 of Ref. [15]). Both M1 and E2 mechanisms lead to similar *P* and *R* branches and thus the E2 and M1 contributions to *P(J)* or *R(J)* lines cannot be separated. The detection of the E2 contribution to the $\nu_2+\nu_3$ band is made possible by the fact that *O* and *S* branches (corresponding to $\Delta J= -2$ and $+2$, respectively) are allowed for E2 transitions but forbidden for M1 transitions. The spectral interval chosen for the present OFCEAS study corresponds to the region of the strongest *S(J)* lines *(J ~10-20)*. Note that the detection of E2 transitions would have been more challenging in the *O*-branch as the *O*-lines fall in a region around 2980 cm$^{-1}$ where stronger E1 lines are located, namely those of the $\nu_2+\nu_3$ E1 band of the $^{16}O^{12}C^{18}O$ minor isotopologue centered near 2982 cm$^{-1}$. This band is indeed allowed for electric-dipole transitions of the non-symmetric isotopologues and, in spite of a natural abundance of about $3.9\times10^{-3}$, the $\nu_2+\nu_3$ E1 band of $^{16}O^{12}C^{18}O$ is stronger than the $\nu_2+\nu_3$ E2 band of the main isotopologue by about two orders of magnitude. (The E1 lines of $CO_2$ as included in the HITRAN database have been included in **Fig. 5** with black and grey circles for the main and minor isotopologues, respectively).

The comparison of our observations to the calculated M1 and E2 lists (**Fig. 5**) leaves no doubt that the S12, S14 and S16 E2 lines are detected. This is the first laboratory detection of electric quadrupole lines in $CO_2$. On average, the observed intensity values of these E2 lines are about 30 % smaller than the *ab initio* values.

Figure 5 shows that the OFCEAS intensity values of the R26-R36 lines agree very well with the empirical intensity values of Ref. [24] relying on FTS measurements. Nevertheless, based on the *ab initio* intensity predictions [14], the intensities of these *R(J)* lines involve a significant E2 contribution and cannot be considered as pure M1 transitions as assumed in Refs. [17,24]. For the set of R26-R36 lines, the E2 contribution based on the *ab initio* E2 intensities represents 21 % of the total line intensity. This percentage is mostly independent of the *R(J)* line because the Hönl-London factors of the considered R26-R36 transitions are close for E2 and M1 transitions ($\frac{J+1}{2}$ and $\frac{J-1}{2}$, respectively [15]). In other words, among the FTS intensity values used in Ref. [15] for the fit of the transition magnetic dipole moment (MDM) squared, those of the *P* and *R* transitions should be reduced by about 21 % keeping unchanged those of the *Q* branch transitions (E2 contribution in the *Q* branch is negligible [15]). Consequently, the amplitude of the fitted value of magnetic dipole moment [18] is believed to be overestimated by about 10 %.

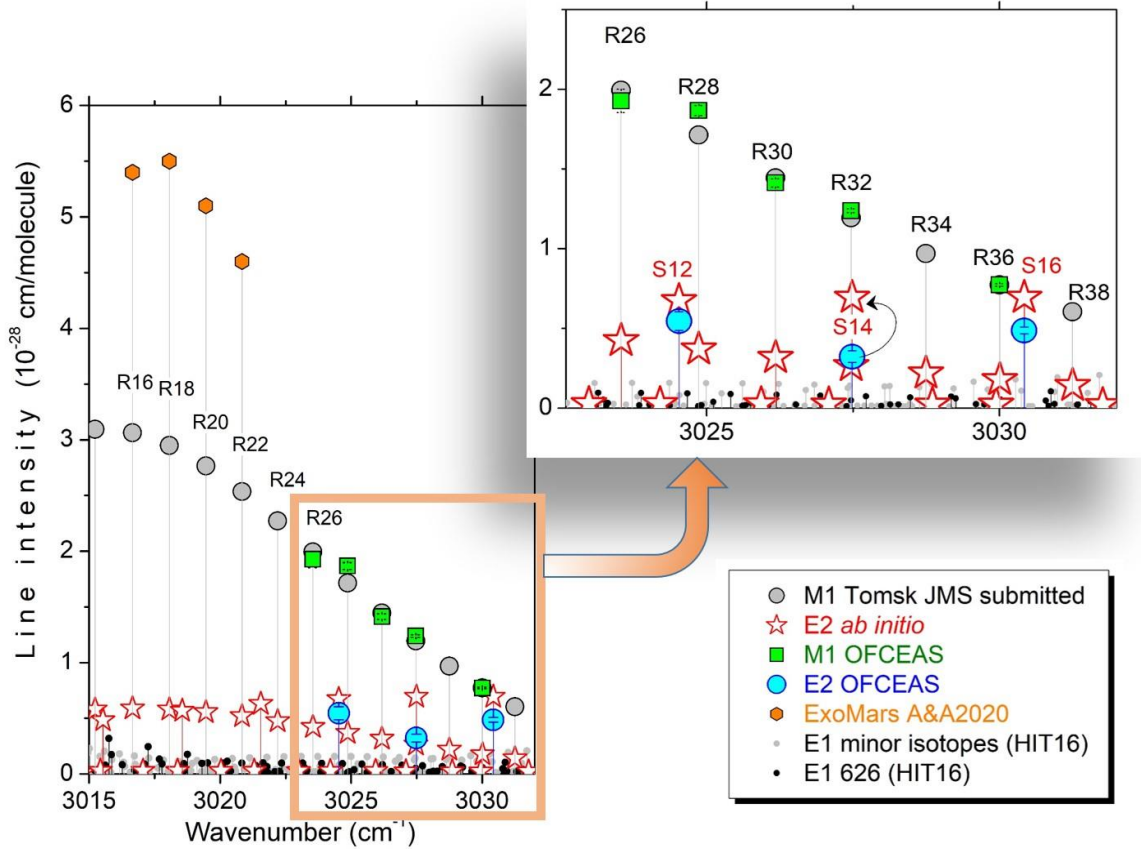

**Fig. 5.**
Overview comparison of the different $CO_2$ transitions contributing to the absorption spectrum in the 3015-3032 cm$^{-1}$ interval:
- HITRAN2016 line list of the electric dipole (E1) transitions of the principal and minor isotopologues, (black and grey circles, respectively),
- electric quadrupole (E2) spectrum computed theoretically (red stars) [14],
- empirical line list of the $\nu_2+\nu_3$ magnetic dipole ("M1") band of $^{12}C^{16}O_2$ (grey circles) [18],
- present OFCEAS measurements of E2 and "M1" transitions (cyan circles and light green squares, respectively). Plotted uncertainties correspond to the statistical error bars given in Table 1. The total uncertainties are larger, on the order of 8 % and 30 % for "M1" and E2 lines, respectively,
- R16-R22 lines measured from the spectrum of the Martian atmosphere in the frame of the ExoMars mission [17] (orange hexagons).

Note that the "M1" intensity values include an E2 contribution (see Text).

As mentioned in the introduction, the first spectrum of the $\nu_2+\nu_3$ M1 band was recorded by the echelle ACS MIR spectrometer onboard the ExoMars Trace Gas Orbiter [16]. Line intensity values of the R16-R22 lines were reported from the analysis of the spectrum of the Martian atmosphere [17]. In order to convert the integrated line absorbances measured on the ExoMars spectrum to absolute intensity values, the nearby $\nu_2+\nu_3$ E1 band of the $^{16}O^{12}C^{18}O$ minor isotopologue was used [17]. The treatment required the knowledge of the Martian isotopic abundance of $^{16}O^{12}C^{18}O$, which was measured and found to coincide with the terrestrial abundance value ($4.01\times10^{-3}$) within 2 % [26]. The ExoMars intensity value of the R16-R22 lines (orange symbols in **Fig. 5**) are overestimated by nearly a factor of two compared to the empirical intensity values of Ref. [24] validated by our present measurements. We have no explanation for this disagreement. In terms of $^{16}O^{12}C^{18}O$ abundance, it would correspond to a Martian abundance twice smaller than the terrestrial value which has to be excluded [26].

4. Conclusion

The first laboratory detection of electric quadrupole transitions of carbon dioxide has been achieved from high sensitivity spectra by Optical-Feedback-Cavity Enhanced Absorption Spectroscopy. The detectivity threshold of the recordings corresponds to an intensity threshold of about $10^{-29}$ cm/molecule which is at the state-of-the-art for the spectral region near 3.3 µm. Recent high sensitivity FTS spectrum with an absorption pathlength of 1058 m [24] reported only "M1" lines in the same region. These FTS spectra provided a laboratory validation of the first observation of the "M1" $\nu_2+\nu_3$ band in the spectrum of the Martian atmosphere by the ExoMars Trace Gas Orbiter [17]. Both the FTS and ExoMars studies considered their observed *P*, *Q* and *R* transitions of the $\nu_2+\nu_3$ band as activated by the magnetic dipole only, while our detection of E2 transitions in the *S* branch validates recent *ab initio* calculations and indicates that the "M1" line intensities reported in Refs. [17,24] actually involve an E2 contribution. Note that the predictions of the E2 *ab initio* spectrum of $^{12}C^{16}O_2$ motivated a detailed examination of the residuals of the best-fit synthetic model of the ExoMars ACS MIR spectrum [14]. Three weak absorption features close to the noise level, with positions (and to a less extend intensities) coinciding with those expected for the S8, S10 and S12 E2 lines were tentatively attributed to the E2 $\nu_2+\nu_3$ band [14] in the residuals of Ref. [17]. This first tentative detection of $CO_2$ E2 transitions in the Martian atmosphere spectrum is thus confirmed by the present OFCEAS laboratory spectra at higher sensitivity.

As regards line intensities, the very good agreement between the OFCEAS and FTS values for the "M1" lines indicate that the intensity values estimated from the Mars spectrum Ref. [17] are overestimated by about a factor of 2. Our E2 intensity values of the S12, S14 and S16 lines are on average 30 % below the predicted *ab initio* values of ref. [14]. This deviation is on the order of our experimental error bars and no uncertainty is provided for the theoretical values. Further experimental measurements, possibly in different spectral regions are thus suitable for validation test of the *ab initio* E2 line intensities in $CO_2$.

By gathering all the available information at disposal, it is possible to propose a recommended line list for the $\nu_2+\nu_3$ band of $^{12}C^{16}O_2$ consisting in: *(i)* the "M1" empirical line list of Ref. [24] with line intensities decreased by 21 % in the *P* and *R* branches for the M1 band and *(ii)* the *ab initio* line list of Ref. [14] for the E2 band.

The "M1" line list of $^{12}C^{16}O_2$ as elaborated in Ref. [24] will be incorporated in the new version of the HITRAN database. The present results about carbon dioxide together with the detection of E2 transitions in different regions of the spectrum of water vapor [12,13] have illustrated the fact that in their present status the spectroscopic databases are not complete for polyatomic molecules, in the sense that they provide lists of electric dipole transitions above an intensity cut-off (*e.g.* $10^{-30}$ cm/molecule) largely below intensity values of the E2 intensities (*e.g.* up to $10^{-26}$ cm/molecule for water vapor [12]). The few experimental tests available for the *ab initio* E2 intensities have validated the theoretical values (for $CO_2$ and water). From our present knowledge, it appears thus suitable to include the *ab initio* E2 line lists of $CO_2$ [14] and water vapor [12] in the next editions of the spectroscopic databases. As regards M1 rovibrational transitions, to the best of our knowledge, no theoretical line list is available for polyatomic molecules and the importance of M1 vibrational bands remain to be investigated.

*Acknowledgments*
*RG received funding from the Agence Nationale de la Recherche (ANR) under grant agreement ANR-18-CE04-0003- 01.*

**References**

1. Gordon IE, et al. The HITRAN2016 Molecular Spectroscopic Database. J Quant Spectrosc Radiat Transf 2017;203:3–69. doi: 10.1016/j.jqsrt.2017.06.038.
2. Jacquinet-Husson N, Armante R, Crépeau N, Chédin A, Scott NA, Boutammine C, et al. The 2015 edition of the GEISA spectroscopic database. J Mol Spectrosc 2016;327:31–72. doi:10.1016/j.jms.2016.06.007
3. Herzberg G. Quadrupole rotation-vibration spectrum of the hydrogen molecule. Nature 1949;163:170.
4. Rothman LS, Goldman. Infrared electric quadrupole transitions of atmospheric oxygen. Applied Optics 1981;20:2182–4.
5. Campargue A, Kassi S, Pachucki K, Komasa J. The absorption spectrum of $H_2$: CRDS measurements of the (2-0) band, review of the literature data and accurate ab initio line list up to 35 000 $cm^{-1}$. Phys Chem Chem Phys 2012;14:802-815. Doi: 10.1039/C1CP22912E
6. Goldman A, Reid J, Rothman LS. Identification of electric quadrupole O2 and N2 lines in the infrared atmospheric absorption spectrum due to the vibration-rotation fundamentals. Geophys Res Lett 1981;8:77-78.
7. Baines H, Mickelson ME, Larson LE, Ferguson DW. The abundances of methane and ortho/para hydrogen on Uranus and Neptune: Implications of New Laboratory 4-0 H2 quadrupole line parameters. Icarus 1995;114, 328–40.
8. Kassi S, Gordon IE, Campargue A. First detection of transitions in the second quadrupole overtone band of nitrogen near 1.44 µm by CW-CRDS with $6\times10^{-13}$ $cm^{-1}$ sensitivity. Chem Phys Lett 2013;582:6–9.
9. Kassi S, Campargue A. Cavity Ring Down Spectroscopy with $5\times10^{-13}$ $cm^{-1}$ sensitivity. J Chem Phys 2012;137:234201.
10. Kassi S, Campargue A. Electric quadrupole and dipole transitions of the first overtone band of HD by CRDS between 1.45 and 1.33µm. J Mol Spectrosc 2011;267:36-42.
11. Vasilchenko S, Mondelain D, Kassi S, Čermák P, Chomet B, Garnache A, Denet S, Lecocq V, Campargue A. The HD spectrum near 2.3 µm by CRDS-VECSEL: electric quadrupole transition and collision-induced absorption. J Mol Spectrosc 2016;326:9–16.doi:10.1016/j.jms.2016.04.002
12. Campargue A, Kassi S, Yachmenev A, Kyuberis AA, Küpper J, Yurchenko SN. Observation of electric-quadrupole infrared transitions in water vapor. Phys Rev Res 2020;2:023091. DOI: 10.1103/PhysRevResearch.2.023091
13. Campargue A, Solodov AM, Solodov AA, Yachmenev A, Yurchenko SN. Detection of electric-quadrupole transitions in water vapour near 5.4 and 2.5 µm. Phys Chem Chem Phys 2020;22:12476. DOI: 10.1039/d0cp01667e
14. Yachmenev A, Campargue A, Yurchenko SN, Küpper J, Tennyson J. Quadrupole transitions in carbon dioxide. PNAS submitted Nov. 2020.
15. Perevalov VI, Trokhimovskiy AYu, Lukashevskaya AA, Korablev OI, Fedorova AF, Montmessin F. Magnetic dipole and electric quadrupole absorption in carbon dioxide. J Quant Spectrosc Radiat Transf (in press). https://doi.org/10.1016/j.jqsrt.2020.107408
16. Korablev O, Montmessin F, Trokhimovskiy A, et al. The Atmospheric Chemistry Suite (ACS) of three spectrometers for the ExoMars 2016 Trace Gas Orbiter. Space Sci Rev 2018;214:7. https://doi.org/10.1007/s11214-017-0437-6
17. Trokhimovskiy A, Perevalov V, Korablev O, Fedorova AF, Olsen K S, Bertaux JL, Patrakeev A, Shakun A, Montmessin F, Lefèvre F, Lukashevskaya A. First observation of the magnetic dipole CO2 absorption band at 3.3 µm in the atmosphere of Mars by ExoMars Trace Gas Orbiter ACS instrument. A&A 2020;639:A142.
18. Morville J, Romanini D, Chenevier M. Patent WO03031949, Université J. Fourier, Grenoble France, 2003.
19. Morville J, Romanini D, Kerstel E. Cavity Enhanced Absorption Spectroscopy with Optical Feedback, in: Cavity-Enhanced Spectroscopy and Sensing, edited by: Gagliardi G and Loock HP, Springer Berlin Heidelberg, 2014;163–209.
20. Lechevallier L, Grilli R, Kerstel E, Romanini D, Chappellaz J. Simultaneous detection of $C_2H_6$, $CH_4$, and $\delta^{13}C$-$CH_4$ using optical feedback cavity-enhanced absorption spectroscopy in the mid-infrared region: towards application for dissolved gas measurements. Atmos Meas Tech 2019;12:3101–9. https://doi.org/10.5194/amt-12-3101-2019.
21. Kerstel E, Iannone RQ, Chenevier M, Kassi S, Jost HJ, Romanini D. A water isotope ( 2 H, 17 O, and 18 O) spectrometer based on optical feedback cavity- enhanced absorption for *in situ* airborne applications. Appl Phys B-Lasers Op- tics 2006;85:397–406. doi: 10.1007/s00340- 006- 2356- 1.
22. Adkins EM, https://doi.org/10.18434/M32200
23. Majcherova Z, Macko P, Romanini D, Perevalov VI, Tashkun SA, Teffo JL, Campargue A. High-sensitivity CW-cavity ringdown spectroscopy of $^{12}CO_2$ near 1.5 µm. J Mol Spectrosc 2005;230:1–21. https://doi.org/10.1016/j.jms.2004.09.011

24. Borkov Yu.G., Solodov AM, Solodov AA, Perevalov VI. Line intensities of the 01111-00001 magnetic dipole absorption band of $^{12}C^{16}O_2$: laboratory measurements. J Mol Spectrosc submitted.
25. Tashkun SA, Perevalov VI, Gamache RR, Lamouroux J. CDSD-296, high-resolution carbon dioxide spectroscopic databank: An update. J Quant Spectrosc Radiat Transfer 2019;228:124–31. doi:10.1016/j.jqsrt.2019.03.001.
26. Krasnopolsky VA, Maillard JP, Owen TC, Toth RA, Smith MD. Oxygen and carbon isotope ratios in the martian atmosphere. Icarus 2007;192:396–403. https://doi.org/10.1016/j.icarus.2007.08.013